Published in *ACS Applied Materials and Interfaces* **2010**, *2*, 2193. doi: 10.1021/am100512c

# Micro-fabrication of Carbon Structures by Pattern Miniaturization in Resorcinol-Formaldehyde Gel

Chandra S. Sharma, Ankur Verma, Manish M. Kulkarni,[#]
Devendra K. Upadhyay and Ashutosh Sharma*

*Department of Chemical Engineering & DST Unit on Nanosciences,*
*Indian Institute of Technology, Kanpur-208016, U.P., India*

## Abstract

A simple and novel method to fabricate and miniaturize surface and sub-surface micro-structures and micro-patterns in glassy carbon is proposed and demonstrated. An aqueous resorcinol-formaldehyde (RF) sol is employed for micro-molding of the master-pattern to be replicated, followed by controlled drying and pyrolysis of the gel to reproduce an isotropically shrunk replica in carbon. The miniaturized version of the master-pattern thus replicated in carbon is about one order of magnitude smaller than original master by repeating three times the above cycle of molding and drying. The micro-fabrication method proposed will greatly enhance the toolbox for a facile fabrication of a variety of Carbon-MEMS and C-microfluidic devices.

* Corresponding author: Ashutosh Sharma. Tel.: +91-512-259 7026; Fax: +91-512-259 0104, e-mail: ashutos@iitk.ac.in.

# Present address: Department of Polymer Engineering, The University of Akron, Akron, OH-44325 (USA).

## INTRODUCTION

Miniaturized and low cost scientific and consumer devices have led to the development of variety of advanced micro- and nanofabrication techniques for patterning of materials (1-27). Carbon is already a widely used material of choice in a variety of micro/nano applications and has even a bigger potential because of its wide window of electro-chemical stability, bio-compatibility, chemical inertness and tailoring of its properties such as porosity and functionality (24-38). Thus, carbon micro/nano structures can be suitable alternatives for sensors, bio-platforms, energy storage devices and other carbon-MEMS applications (24-38). The known microfabrication techniques for glassy carbon (24) include focused ion beam/reactive ion etching, screen printing using carbon based inks (24), and micro-molding followed by pyrolysis (20). Another promising microfabrication technique for carbon is based on direct pyrolysis of patterned photoresists, most notably SU-8 resist (25-33). This technique has attracted much recent attention because of the ease of fabrication with SU-8 and the control of the properties and morphology of the pyrolyzed structures.

In this study, we describe a simple technique for micro-fabrication of glassy carbon structures and surface patterns based on miniaturization of a master pattern by a repeated micro-molding and drying of a resorcinol-formaldehyde (RF) gel, which is also a cheap and effective precursor material to obtain carbon by pyrolysis. The RF gel, introduced first by Pekala (34), remains to be the most widely used precursor for carbon xerogels. An attractive aspect of the RF derived carbon is its promising capacity for being used as an electrode material for energy storage devices (34). As will be shown, the RF gel has a special significance in our context of a suitable micro-fabrication technique for carbon structures because: (a) wet RF gel can be readily micro-molded and removed from the molds of a widely used soft-lithography material PDMS, (b) the RF hydrogels can hold more than 50% (v/v) water and thus their drying under controlled conditions of evaporation result in a significant size reduction by a nearly isotropic and crack-free shrinkage of the gel. The above properties of the RF gel are exploited here to generate patterns in the RF gel starting from a PDMS mold, its replication in RF gel, followed by pattern miniaturization by drying of the gel. The next cycle of miniaturization commences with the shrunk RF gel pattern as a mold for micro-molding of a smaller PDMS replica and repeating the steps of the first cycle. We found that about an order of magnitude reduction in size can thus be obtained after three successive cycles of RF gel drying. The RF gel structure thus fabricated at the end of any cycle can be pyrolyzed to convert the structure into carbon accompanied by some more shrinkage of the structure by the removal of volatile organic material.

Another technique (23) based on reduction of feature size uses projection of a pattern on a photolithography material with the help of a focusing device such as a micro-lens, thus producing a miniaturized surface replica of the original. In contrast, the micro-molding based technique proposed here is not limited to the 2-D surface relief structures and does not involve special equipment such as micro-lens arrays and photolithography—the latter being limited by the wavelength of light. However, suppression of defects requires slow drying conditions making gel-based miniaturization a slow technique. We demonstrate that a wide variety of patterns at different length scales and over large areas could be generated with a single master pattern, while retaining their original shapes. Gel drying and pyrolysis conditions to convert these patterned structures into carbon were optimized to minimize the structural distortions.

## MATERIALS AND METHODS

An epoxy based negative photoresist (SU-8 2015 Microchem, USA) was used for photolithography and Sylgard–184, a two part thermo curable PDMS elastomer (Dow

Chemicals, USA) was employed for making PDMS molds. Resorcinol (99% purity, Qualigens Fine Chemicals, India), Formaldehyde (37% w/v; stabilized by 11-14 wt. % methanol, Loba chemicals, India) and Potassium Carbonate (Loba Chemie, India) were used as received. Double deionized water (Milli-Q) was used for preparing all the samples.

***Fabrication of Master Pattern***

The basic pattern to be replicated and miniaturized was fabricated first in SU-8 photoresist by a maskless photolithography system (38). This patterned photo-resist surface was then used as the first generation master for fabricating a negative-replica in PDMS by micro-molding as described later. In some of the experiments, the aluminum foil from the patterned polycarbonate portion of commercially available optical data storage compact discs (CD) was also used as the master pattern.

***PDMS Mold Preparation (First Generation Pattern)***

A mixture of PDMS elastomer and a curing agent in 10:1 weight ratio was poured on the master pattern and de-aerated in vacuum. The curing was done at 120$^{o}$C for 12 hours. The PDMS negative replica of the first generation master was peeled off gently as shown in Figure 1a. The PDMS replica was exposed to UV-ozone (UVO) to render it hydrophilic and appropriate for use with the aqueous RF gel. This process of replica molding or micro-molding is illustrated in Figure 1. As the RF-sol contains large amount of water, the sol cannot penetrate the capillary spaces completely in a hydrophobic PDMS mold. Thus, making the PDMS stamps hydrophilic by UVO exposure for about 1h or by plasma oxidation at 0.05 torr pressure for about one minute clearly improved the pattern transfer.

***Synthesis of RF Hydrogel***

RF hydrogel was synthesized by the polycondensation of resorcinol and formaldehyde in water (W), in the presence of a basic catalyst-- Potassium carbonate. Resorcinol added to formaldehyde and stirred for 15 min produced a clear yellowish solution. An aqueous solution of potassium carbonate was then added and the mixture was stirred for 30 min at room temperature until the colorless RF sol changed to golden yellow. The resorcinol to formaldehyde molar ratio was 0.50, and resorcinol to water molar ratio was 0.037. Resorcinol to catalyst molar ratio (R/C) was optimized to be 25 for all the cases reported here.

***Pattern transfer from the PDMS stamp to the RF gel surface: Second Generation***

RF sol containing 65%-75% (v/v) water was poured in a container and the patterned, hydrophilic PDMS master was floated on the sol before its gelation point was reached. The sol fills up the patterns because of the capillary force. The sol was then allowed to polymerize to form RF gel and the PDMS stamp was peeled off carefully without distorting the pattern. The patterned RF gel was slowly dried in a controlled environment as mentioned later. This dry gel is called as RF-xerogel and the xerogel surface pattern is referred as the next (second) generation pattern throughout the text.

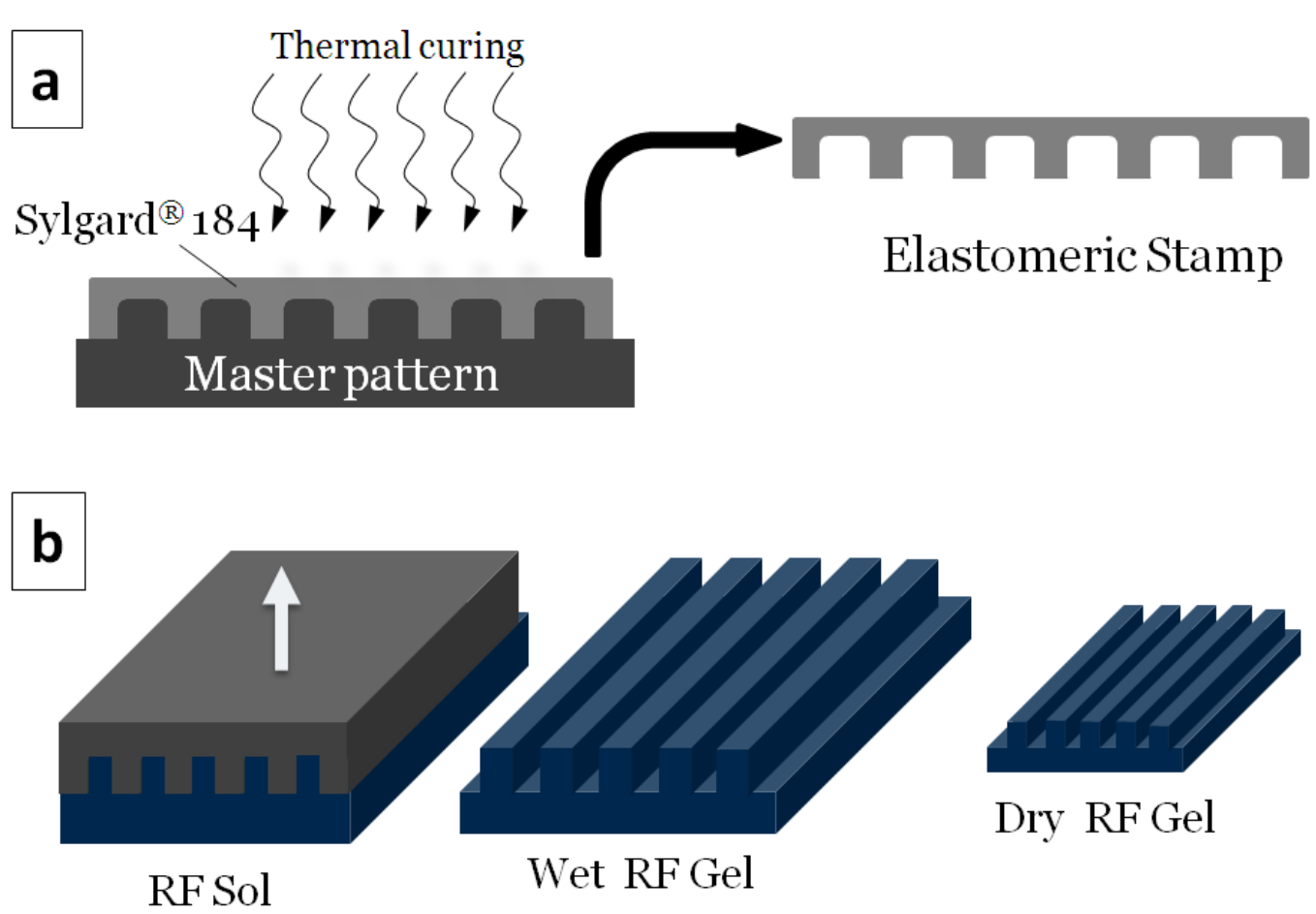

Figure 1: Schematic diagram of (a) replica molding to fabricate elastomeric stamps of sylgard 184 from the master pattern in SU-8, (b) gel patterning by using the elastomeric stamp and shrunk replica of the master in gel after drying

After this complete cycle, micro-molding with Sylgard 184 (PDMS) elastomer was again employed to develop the next generation master from the second generation RF gel pattern as described already in Figure 1b. This step was repeated multiple times in order to generate successive higher generation shrunk patterns on RF-gel surface.

***Pyrolysis of various generation patterns***

The patterned RF xerogels were placed in a quartz boat, and heated to 900°C in a programmed way under nitrogen ($N_2$) atmosphere in a tubular, high temperature furnace for carbonization of the polymer. To obtain carbonaceous surface with good retention of the pattern shape and fidelity, heating conditions were optimized by observing the thermal effects of the dried RF gel using a Thermogravimetric analyzer (Mettler Toledo TGA/SDTA 851e). Once the furnace temperature reached 900$^{o}$C, it was kept constant for 60 min. The furnace was then cooled to room temperature in about 10 h to obtain patterned carbon structures. The inert atmosphere was maintained by purging $N_2$ gas until the furnace attained the room temperature.

## RESULTS AND DISCUSSION

***Optimization of Conditions for Replication in Carbon***

To maintain the monolithicity and to minimize the distortion, buckling and cracks in the RF gels during drying and pyrolysis, various conditions had to be optimized. Some of it was based on the clues from a thermo gravimetric analysis of RF gel as shown in Figure 2. It is observed that there is a significant weight loss (~14%) during initial heating till 120°C. Therefore, patterned RF gels were dried in a closed environment by gradually increasing the temperature from the room temperature to 80°C by 5°C steps in every 12 hours. After 5 days, the patterns were dried at 120°C for 6 hours to ensure complete removal of the water from the gel. Under these conditions the gel sample dried slowly but shrunk uniformly along all the directions without cracking or undergoing undulations and distortions at the surface.

For all the cases reported in this study, diameter to thickness ratio for carbon monoliths was about 6. However, it was observed that the patterns could be transferred and miniaturized successfully with diameter to thickness ratios as high as 8, beyond which a circular RF gel block was found to be more prone to cracking during drying and pyrolysis.

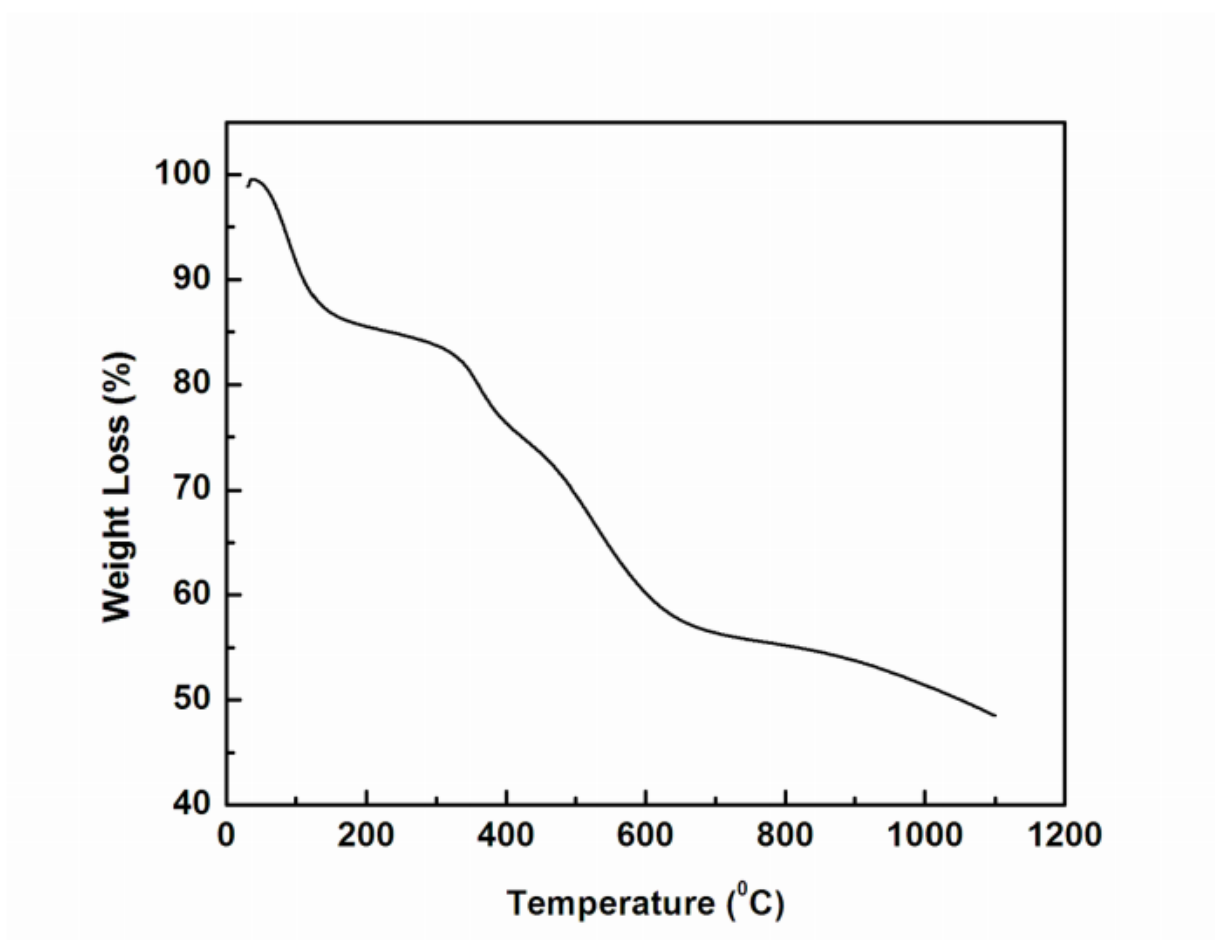

Figure 2: Thermo gravimetric analysis curve of RF hydrogel

For carbonization of the samples, heating rate was also optimized in order to avoid cracking. Based on the TGA results as shown above in Figure 2, it is clear that much of the volume of the gases from RF gel (~25%) escape in the temperature range of 300°C-600°C. Therefore, pyrolysis was done in total of 4 steps. After initial heating rate of 5°C/min from room temperature to 300°C, samples were maintained at 300°C for 1 hour. After that, samples were heated slowly to 900°C with heating rate of 3°C/min and kept at this temperature for 1h before starting natural cooling. Rapid heating may produce cracks in the resulting carbon structures.

The other parameters that play significant roles in allowing a robust pattern transfer are the hydrogel water content and catalyst concentration. In the cases of higher water content than used here, mechanical strength of the RF gel is reduced significantly allowing surface cracks on drying. Higher catalyst concentration, i.e. lower R/C ratio (R/C = 10) results in smaller grains in the RF gel allowing better replication of finer features. However, at the same time, the gel becomes more prone to cracking during drying because of higher crosslinking density, higher internal inter-grain stresses and lower gelation time. Finally, circular shaped gel pieces are less prone to crack on drying as compared to rectangular blocks because of the more uniform stress distribution in the former around its edges. Various types of defects formed outside the optimal processing conditions are shown as supporting information (Figure S1). The patterns miniaturized in this study are over large areas (~ 1-10 $cm^2$). The maximum size patterned successfully without any defects or cracks was 4 cm diameter circular shape of the gel monolith.

***Micro-fabricated and Miniaturized Structures in carbon***

Figure 3 shows variously patterned surfaces of wet RF hydrogel (first column images), dry RF xerogel (second column images) and the carbonized structures (last column images) after three cycles of RF molding, drying and pyrolysis. As observed from Figure 3a-c, line width for the master pattern was 17.5 μm which shrunk to 1.5 ± 0.1 μm after three cycles of miniaturization. Upon pyrolysis, feature size was further reduced to 1.4 ± 0.1 μm. Similarly, size of the pillars as shown in Figure 3d-f, reduced from 43.3 μm (master) to 2.7 ± 0.1 μm after three successive cycles of drying. For hexagonal honeycomb structures, size of the master hexagon patterns used was 122 μm, which after three cycles of miniaturization followed by pyrolysis produced a honeycomb carbon structure with a hexagon size of 8.4 ± 0.3 μm. For complex patterns as shown in Figure 3j-l, size of the logo used was 3.2 mm (only a part of it is visible in the figure) which shrank to 225.2 ± 2.4 μm and 204.7 ± 2.5 μm after three cycles of miniaturization and pyrolysis, respectively.

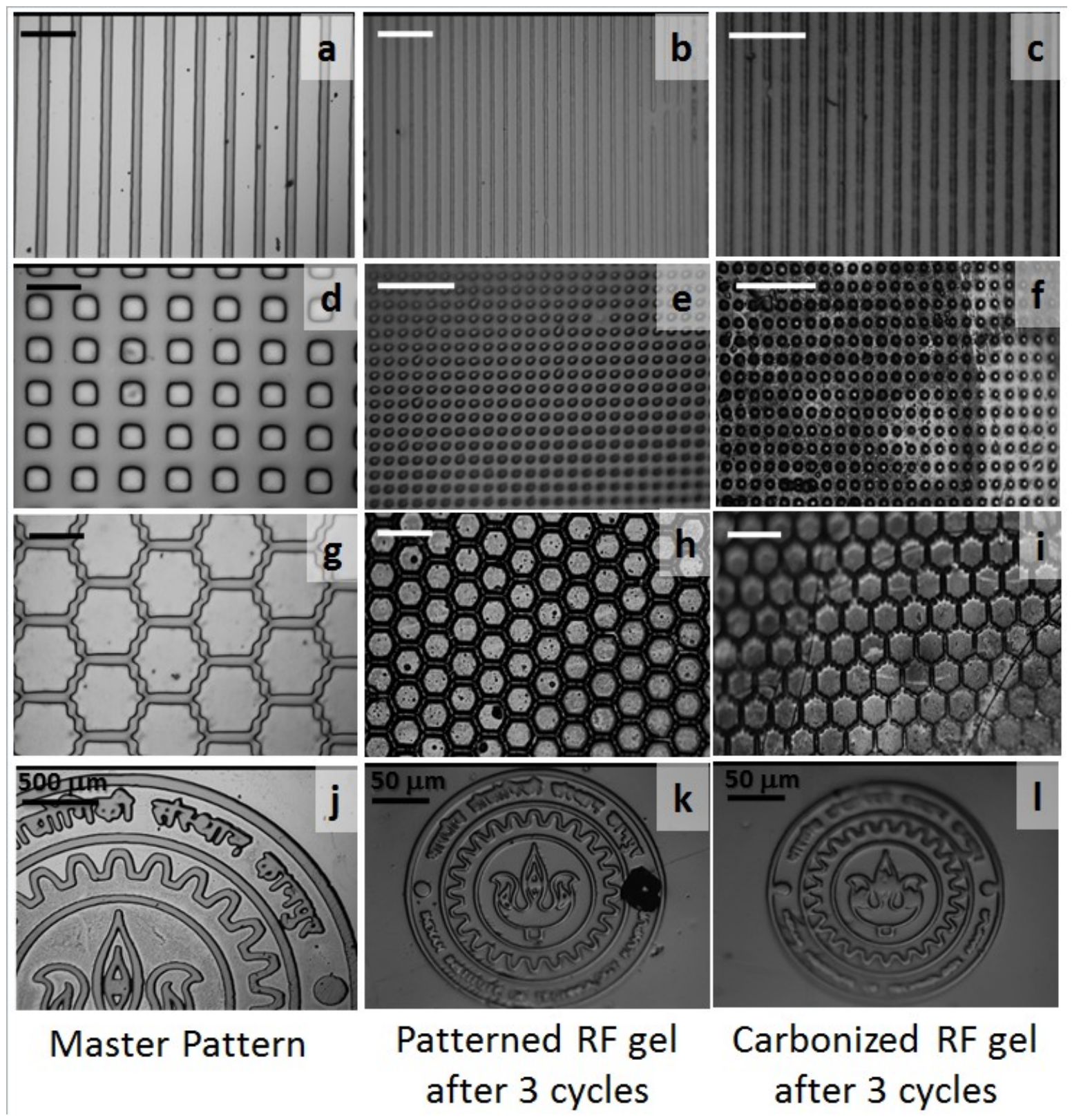

Figure 3: Optical micrographs of master and the RF gel patterns: (a), (b), (c) are open micro channels; (d), (e), (f) are square pillars; (g), (h), (i) are honeycomb structures and (j), (k), (l) are the logo of IIT Kanpur. Black scale bars are 100 μm and white scale bars are 20 μm, unless otherwise noted.

Thus, a wide variety of patterns including microchannels (Figure 3a-c), pillars (Figure 3d-f), and honeycomb structures (Figure 3g-i) to complex patterns (Figure 3j-l) can be transferred and miniaturized using RF-gels and eventually converted to carbon structures. These monoliths can directly be used as patterned electrodes in batteries and super-capacitors after carbonization. However, in applications where carbon structures are to be directly supported on a substrate other than carbon, the PDMS replica of the final cycle can be used as a stamp for micro-contact printing of a suitable carbon precursor (20), including the RF sol employed here. As an example, the array of posts in Figure 3f is a basic architecture useful in microbattery, cell platform and sensor applications. The carbon honeycomb structures such as the one in Figure 3i are reported to be optimal for enhanced transport properties because of its high geometric surface area to volume ratio (39).

Similarly, we could replicate the structural patterns of a CD on to the RF gel surface as shown in Figure 4a-f, confirming the viability of the approach for sub-micron sizes. A nearly isotropically shrunk replica of the master (periodicity ~1.6 μm; height~140 nm, Figure 4b) is obtained in the carbonized structures (periodicity ~750 nm; height ~66 nm) after first cycle (Figure 4d). Further, the second cycle of miniaturization of CD patterns as shown in Figure 4e-f yielded RF patterns with periodicity ~408 nm and feature height ~ 38 nm (periodicity ~379 nm; height ~34 nm for carbonized second generation patterns). However as the feature size reduces further, the wetting of hydrophobic PDMS mold with the hydrogel becomes a limiting factor for further pattern transfer. As discussed earlier, this approach is valid for generating large area patterns with about one order of magnitude reduction in the initial feature size. As shown in Figure 4a, a total of 6 x 4 $cm^2$ area was patterned successfully with the CD pattern. We observe that corner surfaces are slightly bowed in this case, which can be attributed to the rectangular shape of the monolith that led to non-uniform stress distribution across the edges.

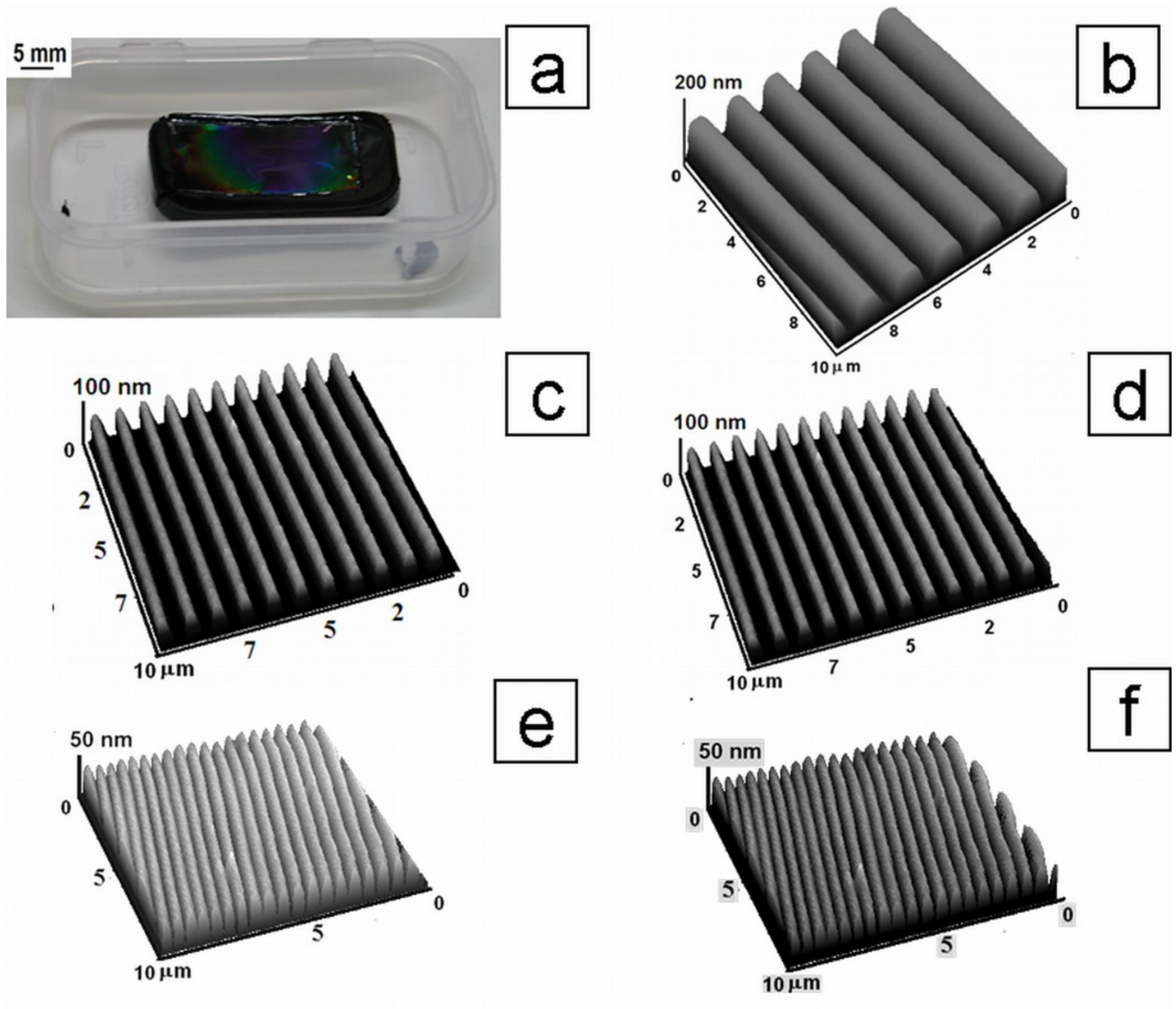

Figure 4: (a) Digital camera image of large area RF-gel block with its surface patterned with aluminum foil of a CD used as a stamp for molding; (b) AFM micrographs showing the original CD pattern (periodicity: ~1.6 µm); (c) first generation dry RF gel pattern (periodicity: ~800 nm); (d) the first generation RF gel derived carbon pattern (periodicity: ~750 nm) (e) second generation dry RF gel pattern (periodicity: ~408 nm), and the second generation RF gel derived carbon pattern (periodicity: ~379 nm)

It is now clear from the Figure 3 and 4 that the RF hydrogel shrinks isotropically during drying, thus allowing miniaturization of earlier generation pattern up to 60% (v/v) in one hydrogel drying cycle. This pattern with reduced size features is subsequently transferred on PDMS surface and further on to the next set of gels and miniaturized, so that the original pattern shrinks by about one order of magnitude in three successive miniaturization cycles as shown in Figure 3.

As observed, the shrinkage reported here for various kinds of patterns is quite reproducible in the tolerance limit of less than 5 %.

Recently, Verma et. al.(40) demonstrated the fabrication of complex 3D micro-channel structures in PDMS by imbedding a template, such as a straight or twisted wire, and its subsequent removal after curing of PDMS. We employed this templating technique combined with the RF gel miniaturization and pyrolysis steps for the fabrication and miniaturization of sub-surface patterns embedded in a carbon block. An example is shown in Figure 5 for the fabrication of circular channels in carbon.

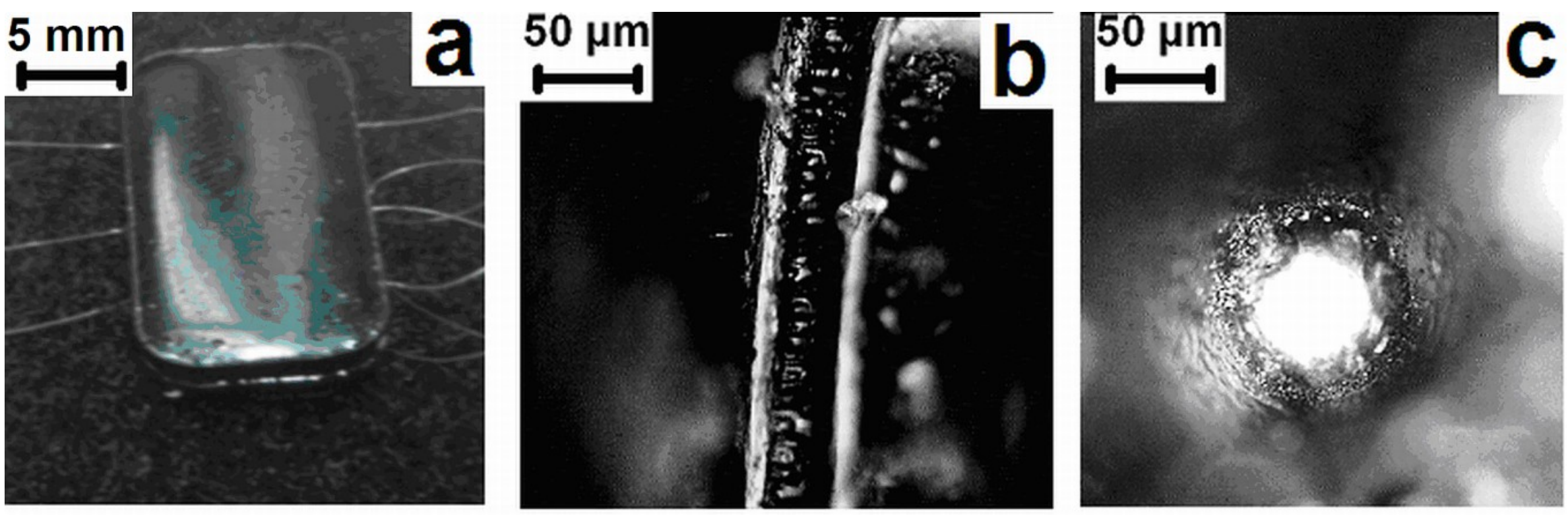

Figure 5: (a) Digital camera image of an RF gel block embedded with nylon threads, (b) lateral view of the carbon micro-channel after template removal, gel shrinkage and pyrolysis, and (c) cross-section of the microchannel.

A 100 micron diameter nylon thread was first embedded in RF sol, which could be removed easily from the wet RF gel before drying. RF gel drying and carbonization produced a 38 micron diameter embedded microchannel in the carbon block. Thus, the proposed technique of miniaturized fabrication in carbon is applicable to 2-D surface patterns, as well as 3-D surface and sub-surface embedded structures.

## Conclusions

In summary, we have demonstrated a novel and facile method for micro-fabrication of a variety of surface patterns and sub-surface micro-structures, such as channels, in carbon by employing micro-molding of an aqueous resorcinol-formaldehyde (RF) gel, followed by its drying and pyrolysis under optimized conditions to ensure isotropic shrinkage and minimal pattern distortion and cracking. Starting from a master pattern in PDMS or other materials, the proposed method replicates the geometry of the master structure on a scale which is about 60% smaller than the original structure, thus enabling miniaturization by about an order of magnitude in three cycles of gel replication and drying. Large area sub-micron surface structures could thus be produced in carbon. This new tool of micro-fabrication in carbon may find a host of applications in Carbon-MEMS and C-microfluidic devices in the area of sensors, bio-platforms, micro-batteries, etc that exploit the electro-chemical and other properties of carbon xerogels which can also be readily tailored for their porosity and functional groups.

## Acknowledgements

This work is supported by the Indo-US Science and Technology Forum, New Delhi, and by an IRHPA grant from the DST. We acknowledge the help of Alok Srivastava, in TGA analysis.

## Supporting Information Available:

Various types of defects developed in RF monoliths under non-optimal drying conditions are given as supporting information.

This material is available free of charge via the Internet at http://pubs.acs.org.